\documentclass[12pt,preprint]{aastex}
\usepackage{times}

\begin{document}

\title{New Infrared Emission Features and Spectral Variations in NGC 7023}
 
\author{
M. W. Werner\altaffilmark{1}, 
K. I. Uchida\altaffilmark{2},
K. Sellgren\altaffilmark{3},
M. Marengo\altaffilmark{4},
K. D. Gordon\altaffilmark{5},
P. W. Morris\altaffilmark{6},
J. R. Houck\altaffilmark{2},
and
J. A. Stansberry\altaffilmark{5}
}

\affil{}
 
\email{
sellgren@astronomy.ohio-state.edu
}

\altaffiltext{1}
{Jet Propulsion Laboratory, MS 264-767, 4800 Oak Grove Drive, Pasadena, 
CA 91109}
\altaffiltext{2}
{Center for Radiophysics and Space Research, Cornell University, Space 
Sciences Building, Ithaca, NY 14853}
\altaffiltext{3}
{Department of Astronomy, Ohio State University, 140 W. 18th Av.,
Columbus, OH 43210}
\altaffiltext{4}
{Harvard-Smithsonian Center for Astrophysics, 60 Garden Street, 
Cambridge, MA 02138}
\altaffiltext{5}
{Steward Observatory, University of Arizona, Tucson, 
933 North Cherry Avenue, AZ 85721}
\altaffiltext{6}
{Spitzer Science Center/Infrared Processing and Analysis Center, 
California Institute of Technology, Pasadena, CA 91125, USA}

\clearpage

\begin{abstract}
We observed the reflection nebula (RN) NGC
7023, with the SH module, and the long-slit SL and LL modules, of the
InfraRed Spectrograph (IRS) on {\it Spitzer}.  We also present
InfraRed Array Camera (IRAC) and Multiband Imaging Photometer for
Spitzer (MIPS) images of NGC 7023 at 3.6, 4.5, 8.0, and 24 $\mu$m.  We
observe the aromatic emission features (AEFs) at 6.2, 7.7, 8.6, 11.3,
and 12.7 $\mu$m, plus a wealth of weaker features.  
We find new unidentified
interstellar emission features at 6.7, 10.1, 15.8, 17.4, and
19.0 $\mu$m.  Possible identifications include aromatic hydrocarbons
or nanoparticles of unknown mineralogy.  
We see variations in relative
feature strengths, central wavelengths, and feature widths,
in the AEFs and weaker emission
features, depending on both distance from the
star and nebular position (SE vs. NW).
\end{abstract}


\section{Introduction}

The interstellar AEFs at 6.2, 7.7, 8.6,
11.3, and 12.7 $\mu$m characterize the mid-infrared (mid-IR) emission
of the diffuse interstellar medium (ISM) of our own and other
star-forming galaxies. \citet{DW81} first suggested that the AEFs were
due to aromatic hydrocarbons. 
\citet{LP84} and
\citet*{ATB85} attributed the AEFs to to polycyclic aromatic
hydrocarbon (PAH) molecules with $\sim$50 carbon atoms.  
We present
here {\it Spitzer} \citep{SIRTF} images and spectroscopy of
\objectname[]{NGC~7023}, a 
RN with AEFs, illuminated by the Herbig Be star
\objectname[]{HD 200775}, at a distance of 430 pc \citep{vdA97}.

\section{Observations}

We imaged NGC 7023 with IRAC \citep{IRAC} in all four channels
(3.6, 4.5, 5.8 and 8.0 $\mu$m) in two epochs (2003 Oct 10 and 2003 Dec
18).  The images were reduced with the Spitzer Science Center (SSC)
IRAC reduction pipeline, and combined with the SSC Mosaicer. This
processing included dark subtraction, flat fielding, mux-bleed
correction, flux calibration, correction of focal plane geometrical
distortions and cosmic ray rejection. Figures \ref{fig_IRAC_2_4} and 
\ref{fig_slits} show our final
images at 3.6, 4.5 and 8.0 $\mu$m.

We obtained images at 24, 70, and 160 $\micron$ with MIPS
\citep{MIPS} on {\it Spitzer}, in the scan map mode at medium rate.  
We covered 10\arcmin\ $\times$ 30\arcmin\ at all three wavelengths.
We reduced the MIPS images 
using the MIPS Instrument Team Data Analysis Tool
\citep{Gordon04}.  
The reduction of the 24 $\micron$ image (Fig. \ref{fig_slits}) followed
preflight expectations, with a 10\% uncertainty on the final absolute
calibration.  
The 70 and 160 $\mu$m images will be
presented in a later paper.

We measured the Short-High (SH) IRS \citep{IRS2} spectrum (9.9 -- 19.4
$\mu$m; $R$ = $\lambda$/$\Delta \lambda$ = 600) of NGC 7023 on 2003
Sept 19.  We obtained Long-Low (LL) long-slit IRS spectra (14.0 --
32.9 $\mu$m; $R$ = 60 -- 130) at two positions on 2003 Oct 1.  These
observations occurred during the checkout phase of {\it Spitzer},
resulting in larger than normal uncertainties in the SH and LL
absolute positions of $\pm$5\arcsec.  We observed the Short-Low (SL)
long-slit IRS spectrum (5.1 -- 15.1 $\mu$m; $R$ = 60 -- 130) of NGC
7023 on 2003 Oct 25.  IRS was sufficiently checked out by this point,
resulting in a much smaller absolute positioning for the SL data
of $\pm$1\arcsec.

The data were reduced and spectra extracted using Cornell IRS
Spectroscopy Modeling Analysis and Reduction Tool \citep{Higdon04}.  
The spectra were calibrated by applying the spectrum and model
template of point source calibrator $\alpha$ Lac to the intermediate
unflat-fielded product of the SSC pipeline.  We subtracted blank sky
spectra only from the SH spectrum, observed as {\it Spitzer} cooled,
to remove thermal emission from the then warm ($\sim$45 K) baffles of
the telescope.  All spectra were extracted from fixed pixel ranges
over the slit, except for SH where we extracted the spectrum from over
the entire aperture.

Figure \ref{fig_slits} shows that the SL slit intersects the filaments
seen to the NW of HD 200775.  Our primary LL slit intersects both the
NW filaments and HD 200775 (Fig. \ref{fig_slits}).  The SL slit
crosses our secondary LL slit at Position A, shown on Figure
\ref{fig_slits}.  We obtained the SH spectrum at Position B, also
marked on Figure \ref{fig_slits}.

\section{Results}

We present our spectra of NGC 7023 in Figures \ref{fig_slll_sh},
\ref{fig_sl}, and \ref{fig_ll}.  Table 1 gives the 
central wavelengths and
nebular positions at which we discovered new interstellar, spectrally resolved,
emission features.

\begin{center}
\begin{tabular}{rr}
\multicolumn{2}{c}{{\bf TABLE 1}}\\
\multicolumn{2}{c}{{\bf New Infrared Emission Features}}\\
\multicolumn{1}{c}{$\lambda_c$}&
\multicolumn{1}{r}{Positions}\\
\multicolumn{1}{c}{($\mu$m)}&
\multicolumn{1}{r}{Detected$^{\it a}$} \\
6.7    &  S1--3 \\     
10.1   &  B, S1--3 \\ 
15.8    &  A, L6--7 \\
17.4    &  A, B, L4--6\\ 
19.0    &  L1--5 \\   
\end{tabular}
\end{center}

\noindent
($a$) Positions where features are detected.  
The nebular offset from HD 200775 of each position
is labeled on Figure \ref{fig_slll_sh} (Positions A, B),
Figure \ref{fig_sl}
(Positions S1 -- S8)
and Figure \ref{fig_ll}
(Positions L1 -- L7).

Figure \ref{fig_slll_sh}a shows the combined SL and LL spectrum (5 --
33 $\mu$m) at Position A.  The absolute flux calibrations of SL1 and
SL2 are uncertain by 25\%, yet the relative agreement between them is
much better than this factor.  The SL and LL spectra were not expected to
match and we had to divide the LL2 and LL1 spectra by factors of
$\sim$2 -- 3, after correction for beam size, to get all spectral
segments to match.  We emphasize that this composite spectrum of Position A
shows qualitatively the spectral features observed in this nebular
region, and should {\it not} be used for quantitative analysis.  The
brightest spectral emission features seen in the spectrum are the AEFs
at 6.2, 7.7, 8.6, 11.3, and 12.7 $\mu$m.  Weaker emission features are
also present, both spectrally resolved emission features and
unresolved H$_2$ emission lines.

Figure \ref{fig_slll_sh}b shows our SH spectrum (9.9 -- 19.4 $\mu$m)
at Position B.  \citet{Moutou00} and \citet{vanKerckhoven00} have
observed this wavelength region in NGC 7023 at a similar resolution
but a different spatial position, using the SWS spectrometer on {\it
ISO}, at a lower signal-to-noise.  Position A is near the
H$_2$-emitting NW filament, while Position B is in a region with
little H$_2$ emission \citep{Lemaire96}.  These two positions sample
different physical conditions, yet show similar spectra.  The higher
spectral resolution of our SH spectrum (Fig. \ref{fig_slll_sh}b)
allows us to easily distinguish between unresolved emission lines,
such as the 0 -- 0 S(1) H$_2$ line at 17.03 $\mu$m, and resolved
emission features, such as the adjacent 17.4 $\mu$m feature.  Weak
emission features that are barely spectrally resolved in Figure
\ref{fig_slll_sh}a are clearly resolved in Figure \ref{fig_slll_sh}b. 

One of the great strengths of {\it Spitzer} is the multiplex
advantage, and the ability to discern spectral and spatial
differences, of the long-slit spectrographs SL and LL.  In Figure
\ref{fig_sl}, we present the results from our long-slit SL
observations.  The AEFs are the brightest spectral features at all
positions.  We detect, in SL spectra closest to the star (S1 -- S4),
the split between the 7.6 and 7.8 $\mu$m features that mainly comprise
the ``7.7'' $\mu$m AEF (Fig. \ref{fig_sl}).

The 11.0 $\mu$m feature is clearly separated from the 11.3 $\mu$m AEF
in Figure \ref{fig_slll_sh}b.  In Figure \ref{fig_sl}, we can observe
the 11.0 $\mu$m feature blended with the 11.3 $\mu$m AEF, in SL
spectra closest to the star (S1 -- S3).  This blend, and its spatial
variations, causes the 11.3 $\mu$m AEF to appear to decrease in FWHM
and shift to longer wavelengths with increasing distance, $d_*$, from
the star.

Figure \ref{fig_ll} illustrates our LL2 spectra (14.0 -- 21.1 $\mu$m)
along the long-slit of IRS-LL.  Figure \ref{fig_ll},
together with Figure \ref{fig_slll_sh}, demonstrate that longward of
the five bright AEFs, there continues to be complex spectral
structure.  Most prominent in Figure \ref{fig_ll} is an emission
feature at 16.4 $\mu$m, NW of the star (L4 -- L7).  Fig. \ref{fig_ll}
also illustrates emission features NW of the star, at 15.8 $\mu$m,
17.4 $\mu$m, and 17.8 $\mu$m.  These features to the NW are confirmed
by the SL+LL spectrum of Position A (Fig. \ref{fig_slll_sh}a) and the
SH spectrum of Position B (Fig. \ref{fig_slll_sh}b).  Figure
\ref{fig_ll} also shows an emission feature both SE and NW of the star
at 19.0 $\mu$m.  Figures \ref{fig_slll_sh}, \ref{fig_sl}, and
\ref{fig_ll} yield abundant evidence of pure rotational lines of
H$_2$, marked on Figures \ref{fig_sl} and \ref{fig_ll}.
\citet{Fuente99, Fuente00} have previously studied pure rotational
lines of H$_2$ in NGC 7023.

Figures \ref{fig_slll_sh}, \ref{fig_sl}, and \ref{fig_ll} all show a
faint but non-zero continuum, observed at 5 -- 20 $\mu$m, in addition
to spectral features and lines.  This continuum emission is also
detected spectroscopically at 2 -- 4 $\mu$m in NGC 7023
(\citealt*{SWD83}; \citealt*{Martini99}), along with a strong 3.3 $\mu$m
AEF and plateau of emission at 3.2 -- 3.6 $\mu$m.  These spectra
demonstrate clearly that the emission in the IRAC 3.6 $\mu$m filter
(3.2 -- 3.9 $\mu$m) in NGC 7023 (Fig. \ref{fig_IRAC_2_4}) is due to a mix
of this continuum emission with the 3.3 $\mu$m AEF and its
accompanying 3.2 -- 3.6 $\mu$m plateau emission.

\citet{SWD83} and \citet{Sellgren84} proposed that the 2 -- 5 $\mu$m
continuum in NGC 7023 and other similar RN is due to non-equilibrium
thermal emission from tiny grains ($\sim$1 nm), stochastically heated
to high temperatures ($\sim$1000 K) for a brief time by single stellar
photons.  The IRAC 4.5 $\mu$m filter covers 4.0 -- 5.0 $\mu$m, a
region in which no significant PAH features have been identified.  We
believe, therefore, that the emission within the IRAC 4.5 $\mu$m
filter is completely due to this tiny grain continuum emission.  The
IRAC 8.0 $\mu$m filter covers 6.5 -- 9.3 $\mu$m.  Figures
\ref{fig_slll_sh}a and \ref{fig_sl} clearly demonstrate that the IRAC
8.0 $\mu$m filter is dominated by AEF emission at 7.7 and 8.6
$\mu$m.

The 20 -- 33 $\mu$m spectrum of Position A (Fig. \ref{fig_slll_sh}a)
shows a strong rise to longer wavelengths, producing the emission in
the 24 $\mu$m MIPS images (Fig. \ref{fig_slits}).  {\it IRAS} 25
$\mu$m observations of the RN 23 Tau \citep*{Castelaz87} agree with
predictions \citep{DA85} that emission in this wavelength region is
due to stochastically heated tiny grains.  We expect this to be true
of NGC 7023 as well.

\section{Discussion}

\subsection{Discoveries of new emission features}

We have discovered new ISM emission features at 6.7, 10.1, 15.8, 17.4,
and 19.0 $\mu$m in NGC 7023.  These features are spectrally resolved,
so are not atomic or molecular emission lines.  All five new features
are observed in multiple positions with SH, SL or LL, giving
confidence that they are not instrumental artifacts.  The 10.1 $\mu$m
feature seen in SL, 
and the 17.4 $\mu$m feature seen in LL,
are both clearly detected in SH. 
These ISM
spectral emission features are among the first spectroscopic
discoveries made by {\it Spitzer}.

Currently, these features have no identification.  Possible
identifications could include PAH bands, such as C--H out-of-plane
bending modes for 10.1 $\mu$m \citep{Hony01}, or C--C--C bending modes
for 15.8, 17.4, and 19.0 $\mu$m \citep{vanKerckhoven00}.  The spectral
structure at 6 -- 9 $\mu$m is quite complicated, with only tentative
identifications for well-known features \citep{Peeters02}, making it
difficult to predict whether the 6.7 $\mu$m feature might fit into the
PAH model.  Alternate possibilities, particularly at the longer
wavelengths, include nanoparticles ($\sim$ 1 nm) of a specific mineral
composition.  For instance, \citet{Molster01} identify a mixture of
PAH species between 3 and 12 $\mu$m, and various crystalline and
amorphous silicates beyond 18 $\mu$m, in the spectrum of the planetary
nebula \objectname[]{NGC 6302}.  Other mineral species, such as simple
metal oxides, or other combinations of abundant refractory materials,
also remain to be explored.

\subsection{Spatial variations in spectral features}

We observe marked changes in spectral features across NGC 7023.  The
nebula SE of the star is where the new 19.0 $\mu$m feature is brightest.
To the NW of HD 200775, the 19.0 $\mu$m feature fades and the 16.4
$\mu$m feature becomes the brightest feature in our LL2 spectra.

Our long-slit observations show that the relative strengths of various
pairs of emission features vary with $d_*$.  We consider feature pairs
observed in the same order and module of IRS and find that the ratios 7.8
$\mu$m/7.6 $\mu$m, 7.4 $\mu$m/7.6 $\mu$m, 11.3 $\mu$m/11.0 $\mu$m, and
11.3 $\mu$m/7.7 $\mu$m increase with increasing $d_*$.

One of the most marked spectral changes is an apparent weakening of
the 8.6 $\mu$m AEF with increasing $d_*$.  \citet{Uchida00} and
\citet{Cesarsky00} observe this same phenomenon in other RN, and
attribute it to a broadening of the 7.7 $\mu$m AEF with increasing
$d_*$.  We confirm that the 7.7 $\mu$m AEF is broader, and also find
that its central wavelength increases with increasing $d_*$.  This
suggests that the wing of the 7.8 $\mu$m feature, as it grows in
strength relative to the 7.6 $\mu$m feature, steadily overwhelms the
8.6 $\mu$m AEF until it is barely visible.  The 6.2 $\mu$m AEF, like
the 7.7 $\mu$m AEF, appears broader at larger $d_*$.  The equivalent
width of the 6.2 $\mu$m AEF also markedly decreases with increasing
$d_*$.

Another striking spectral variation with $d_*$ is the changes in the
11 -- 14 $\mu$m region.  Close to the star, distinct features at 11.0,
11.3, 12.0, and 12.7 $\mu$m can be distinguished.  Far from the star,
this spectral region can only be fit by a blend of the 11.3 $\mu$m AEF
and a broad bump of emission, centered at 12.5 $\mu$m and having a
FWHM of 2.0 $\mu$m.  This broad 12.5 $\mu$m feature increases with
increasing $d_*$, and is not observed close to the star.  The same 11
-- 14 $\mu$m behavior with increasing $d_*$ has been observed in the
RN Ced 201 \citep{Cesarsky00}.

\subsection{Imaging}

In the inner regions of the nebula, we find strong similarities
between the IRAC 4.5 $\mu$m continuum image and the IRAC 8.0 $\mu$m
AEF image.  The ring of emission between HD
200775 and the NW filaments is real, and is observed in ground-based
images of the 3.3 $\mu$m AEF \citep{An03}.  

On the largest scales ($\sim$6\arcmin, or $\sim$0.8 pc), our 4.5, 8.0,
and 24 $\mu$m images 
show an hour-glass shape, containing little IR emission, which is
clearly outlined by a narrow rim of IR emission.  The filaments within
$\sim$1\arcmin\ of HD 200775 define the narrow waist of the hourglass
shape.  This spatial structure has been observed before, in 1--0
$^{13}$CO, 2--1 $^{12}$CO, and 3--2 $^{13}$CO
\citep{Gerin98,Fuente98}, at 10 -- 20\arcsec\ resolution.
\citet{Watt86} and \citet{Fuente98} have argued that the hourglass 
shape is the fossil remnant of a bipolar outflow from the Herbig Be
star HD 200775, implying that the large-scale mid-IR emission
outlines the walls of the cavity produced by the
earlier outflow.

{\it Note added in manuscript.}---After this paper was completed,
M. Jura of UCLA kindly pointed out to us that the positions and
widths of the new 17.4 and 19.0 $\mu$m features agree rather well
with those of the two lower frequency transitions of the
C$_{60}$ molecule \citep{Frum91}.  It is not possile to establish
detection of interstellar C$_{60}$ on the basis of the data in
the present paper, but we will explore this potential identification
with additional observations and analysis.  Note, however, that
\citet{Moutou99} have set limits on the abundance of
C${60}$ and C$_{60}^+$ in NGC 7023 based on nondetection of the
two higher frequency vibrational transitions between 7 and 9 $\mu$m.

\acknowledgements
We thank the people whose dedicated work made the Spitzer Space
Telescope a reality (see \citealt{SIRTF}).  This work is based on
observations made with the Spitzer Space Telescope, which is operated
by the Jet Propulsion Laboratory, California Institute of Technology
under NASA contract 1407. Support for this work was provided by NASA
through Contract Number 1257184 issued by JPL/Caltech.  KS thanks the
JPL Center for Long-Wavelength Astrophysics for support.

\clearpage
\begin{figure*}
\figurenum{1}
\caption{IRAC images of the central 6\arcmin\ $\times$ 6\arcmin\ of
the RN NGC 7023, at 4.5 $\mu$m {\it (top)} and 8.0 $\mu$m {\it
(bottom)}.  NGC 7023 spectra show that the IRAC 4.5~$\mu$m filter is
continuum emission and the IRAC 8.0~$\mu$m filter is primarily AEF
emission.  Both observations were made with the 12 s High Dynamic
Range mode, consisting of a 1.2 s exposure for observing bright field
stars, followed by a 10.4 s exposure for detecting the low surface
brightness nebular emission.  Even in this mode, the central star,
HD 200775, is saturated in both images. North is up; east is to the left.}
\label{fig_IRAC_2_4}
\end{figure*}

\begin{figure*}
\figurenum{2}
\caption{MIPS image of the central $\sim$12\arcmin\ $\times$
$\sim$10\arcmin\ of NGC 7023, at 24 $\mu$m {\it (top)}, and IRAC image
of the central $\sim$3\arcmin\ $\times$ $\sim$3\arcmin\ region,
at 3.6 $\mu$m {\it (bottom)}.  We illustrate how the IRS slits overlay
different parts of the central
regions of our 3.6 $\mu$m IRAC image, which is our highest spatial
resolution {\it Spitzer} image.  We extracted and combined SL2 and SL1
spectra at eight spatial locations.  Spectrum S1 (42\arcsec\ W
8\arcsec\ N) is closest to the star and spectrum S8 (38\arcsec\ W
58\arcsec\ N) is farthest from the star.  We extracted LL2 spectra at
seven spatial positions.  Offsets range from 25\arcsec\ E 19\arcsec\ S
(spectrum L1) to 49\arcsec\ W 36\arcsec\ N (spectrum L7).  Spectrum L3
is HD 200775.  Position A (the intersection of SL and LL) and Position
B (the slit location for SH) are marked.  North is up; east is to the left.}
\label{fig_slits}
\end{figure*}
 
\begin{figure*}
\figurenum{3}
\caption{
{\it (a)}
Combined 5 -- 33 $\mu$m normalized spectrum ($R$ = 60 -- 130) of
Position A in NGC 7023 {\it (top)}.  We combined SL2 (5.1 -- 7.6
$\mu$m; {\it red}), SL1 (7.5 -- 15.1 $\mu$m; {\it green}), LL2 (14.0
-- 21.1 $\mu$m; {\it magenta}), and LL1 (20.9 -- 32.9 $\mu$m; {\it
blue}) spectra.  No scaling was done between SL2 and SL1.  LL2 and LL1
were multiplied by factors of order $\sim$2--3, after scaling by the
beam size, in order to match SL1.
{\it (b)}
Normalized SH spectrum (9.9 -- 19.4 $\mu$m; $R$ = 600) of Position B
in NGC 7023 {\it (bottom)}, extracted from the entire entrance slit
(4\farcs7 $\times$ 11\farcs3).  The spectrum is a composite of eleven
orders, with substantial overlap between each.  Relative uncertainty
in the flux calibration between orders is $\sim$7\%.  Adjacent orders
are shown in contrasting colors.}
\label{fig_slll_sh}
\end{figure*}

\begin{figure*}
\figurenum{4}
\caption{Long-slit SL ($R$ = 60 -- 130) spectra of NGC 7023.  We
extracted and combined SL2 (second order; 5.1 to 7.6 $\mu$m) and SL1
(first order; 7.5 to 15.1 $\mu$m) spectra at eight spatial locations,
with a 7\farcs2 $\times$ 3\farcs6 box.  No scaling was done; SL1 and
SL2 agree well at overlapping wavelengths.  The absolute flux
uncertainty in SL is 25\%.  The $y-$axis is flux density (Jy).  The
panel for each of the spectra S1 {\it (bottom)}, S2, S3, S4, S5, S6,
S7, and S8 {\it (top)} is labeled with the offset of the spectrum,
from HD 200775, in arcsec, and the wavelengths of the H$_2$ lines 0--0
S(5) 6.91 $\mu$m, 0--0 S(3) 9.66 $\mu$m, and 0--0 S(2) 12.28 $\mu$m
are marked {\it (vertical lines)}.
}
\label{fig_sl}
\end{figure*}

\begin{figure*}
\figurenum{5}
\caption{Long-slit LL2 (14.0 -- 21.1 $\mu$m; $R$ = 60 -- 130) spectra
of NGC 7023, extracted at seven spatial positions, each with a
15\farcs3 $\times$ 10\farcs6 box.  The $y-$axis is flux density (Jy).
The absolute flux uncertainty in LL2 is 25\%.  The panel for each of
the spectra L1 {\it (bottom)}, L2, L3, L4, L5, L6, and L7 {\it (top)}
is labeled with the offset of the spectrum, from HD 200775, in arcsec,
and the wavelength of the H$_2$ line 0--0 S(1) 17.03 $\mu$m is marked
{\it (vertical line)}.  Spectrum L3 is HD 200775.  }
\label{fig_ll}
\end{figure*}

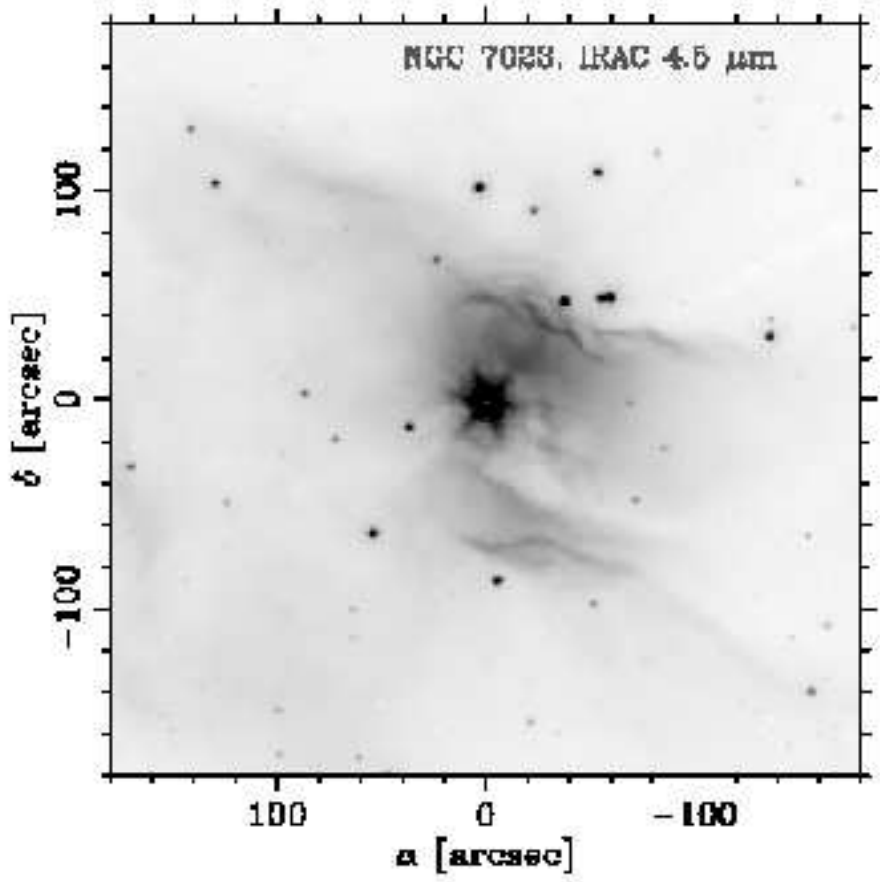
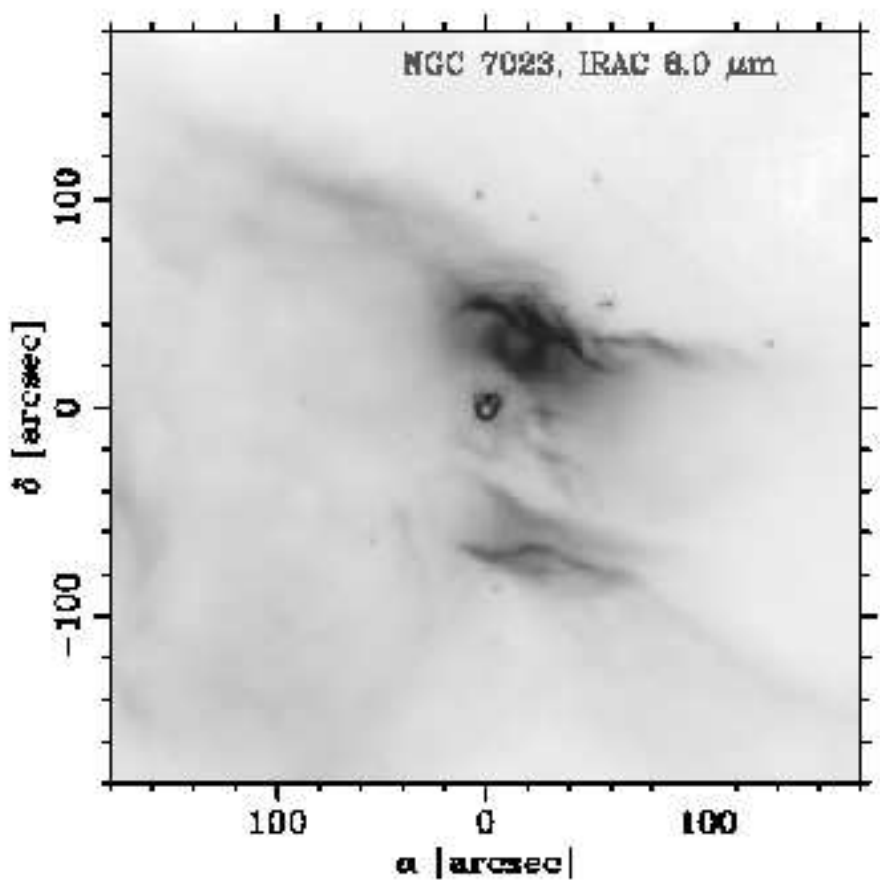
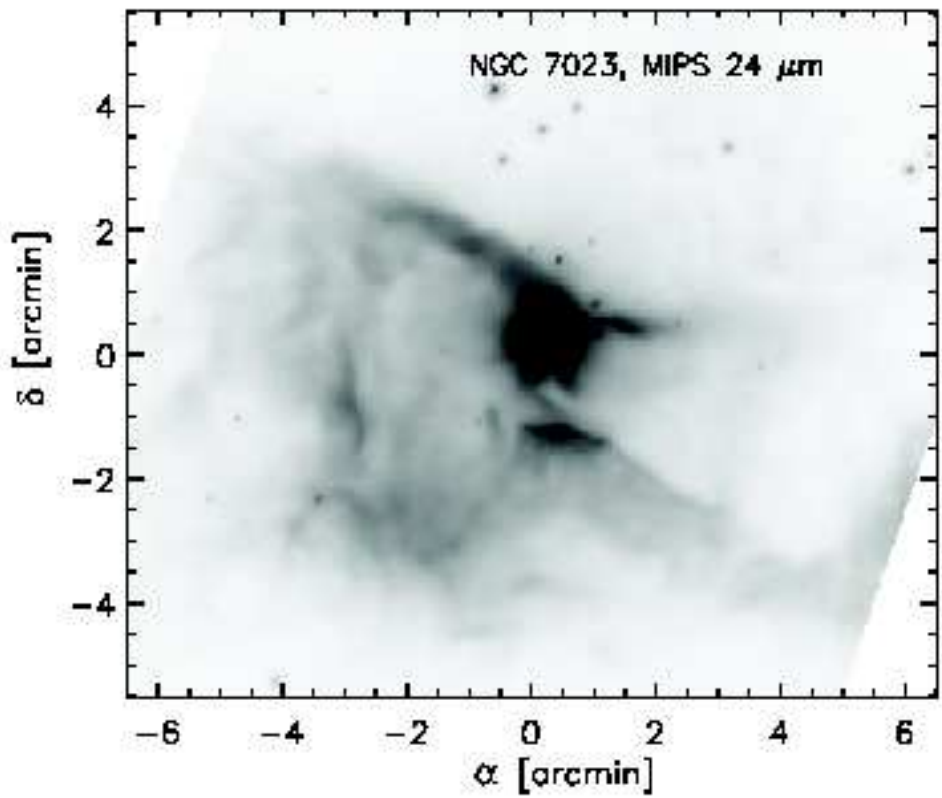
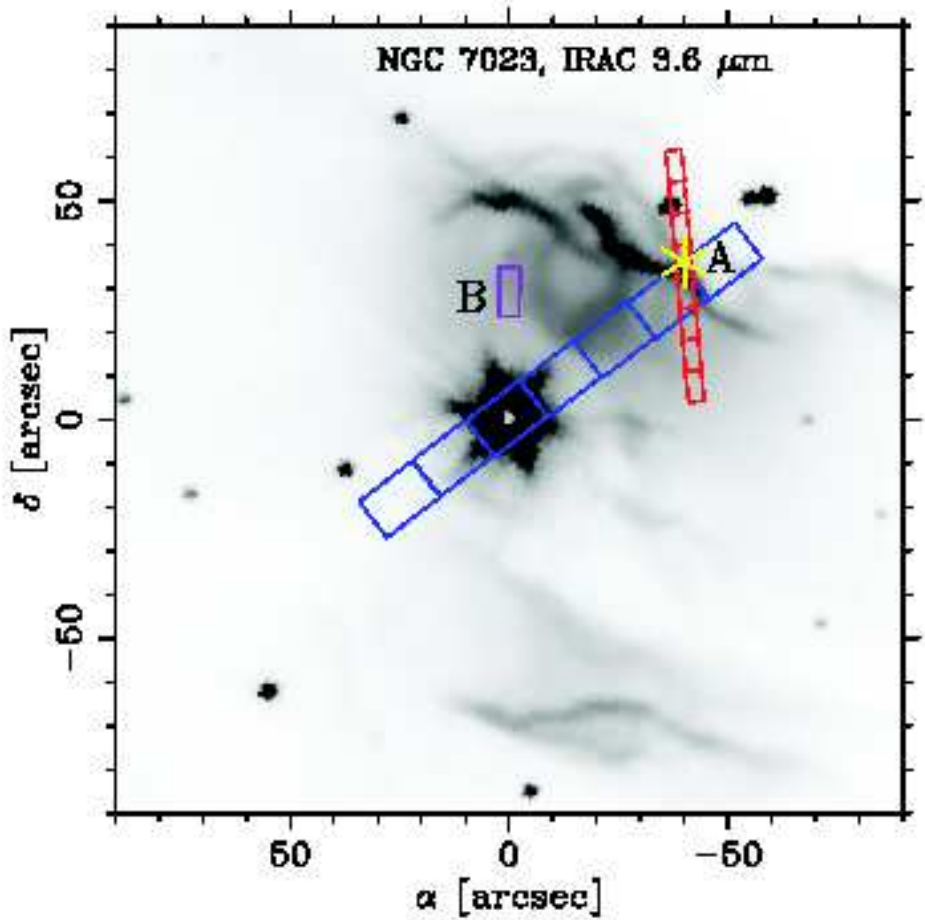
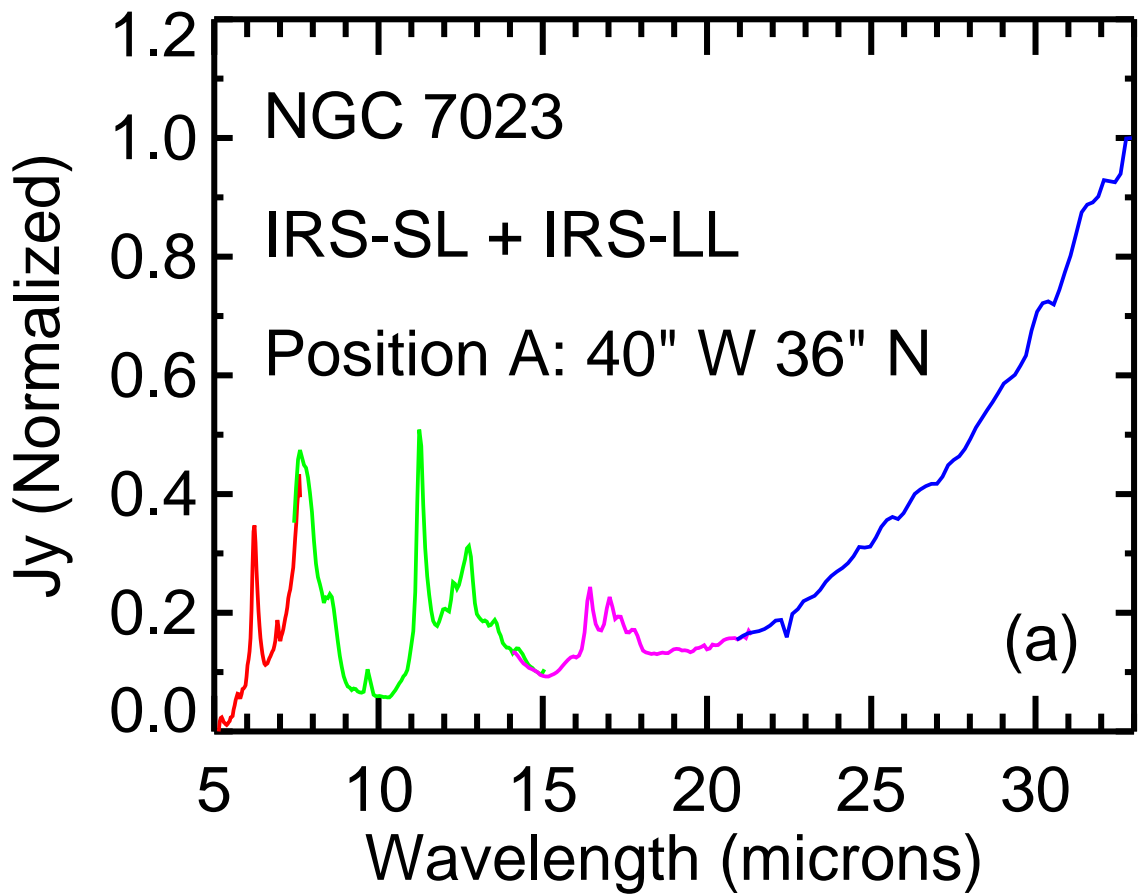
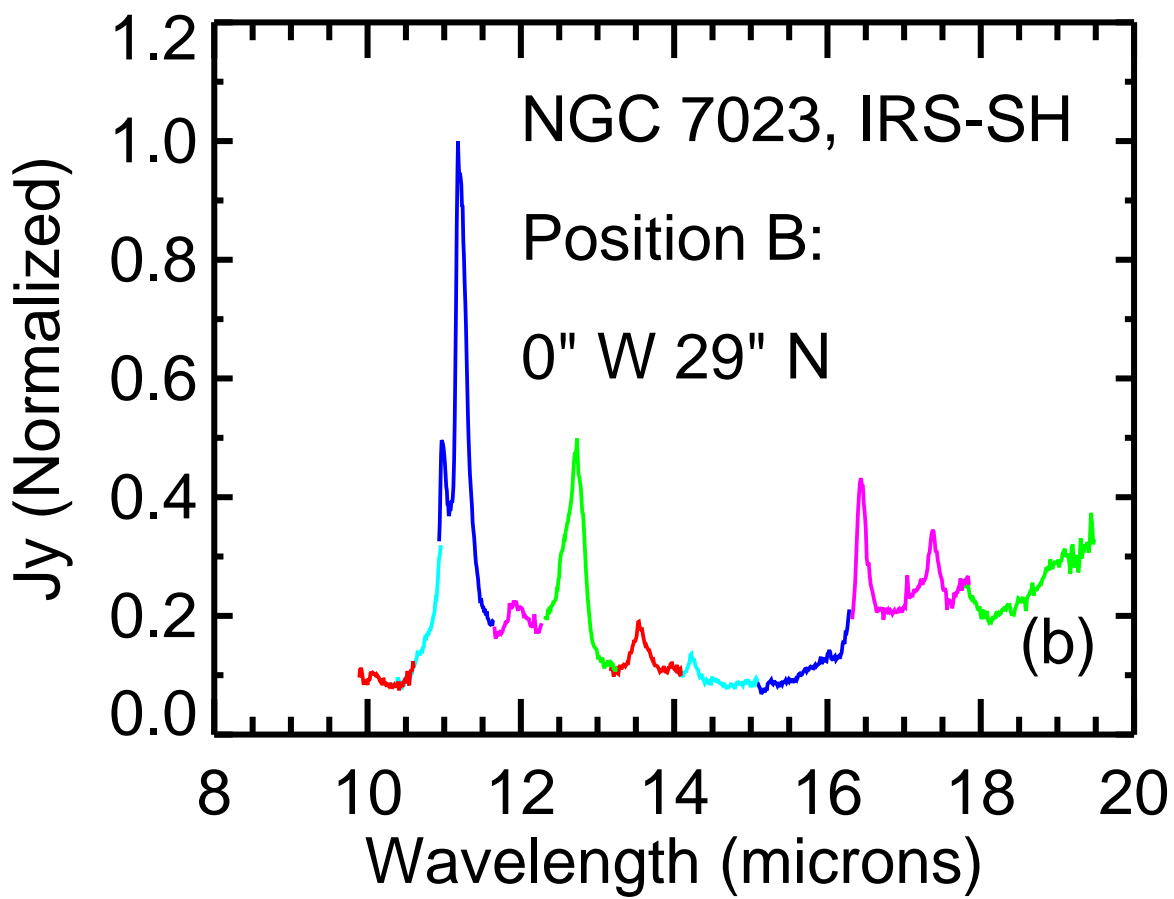
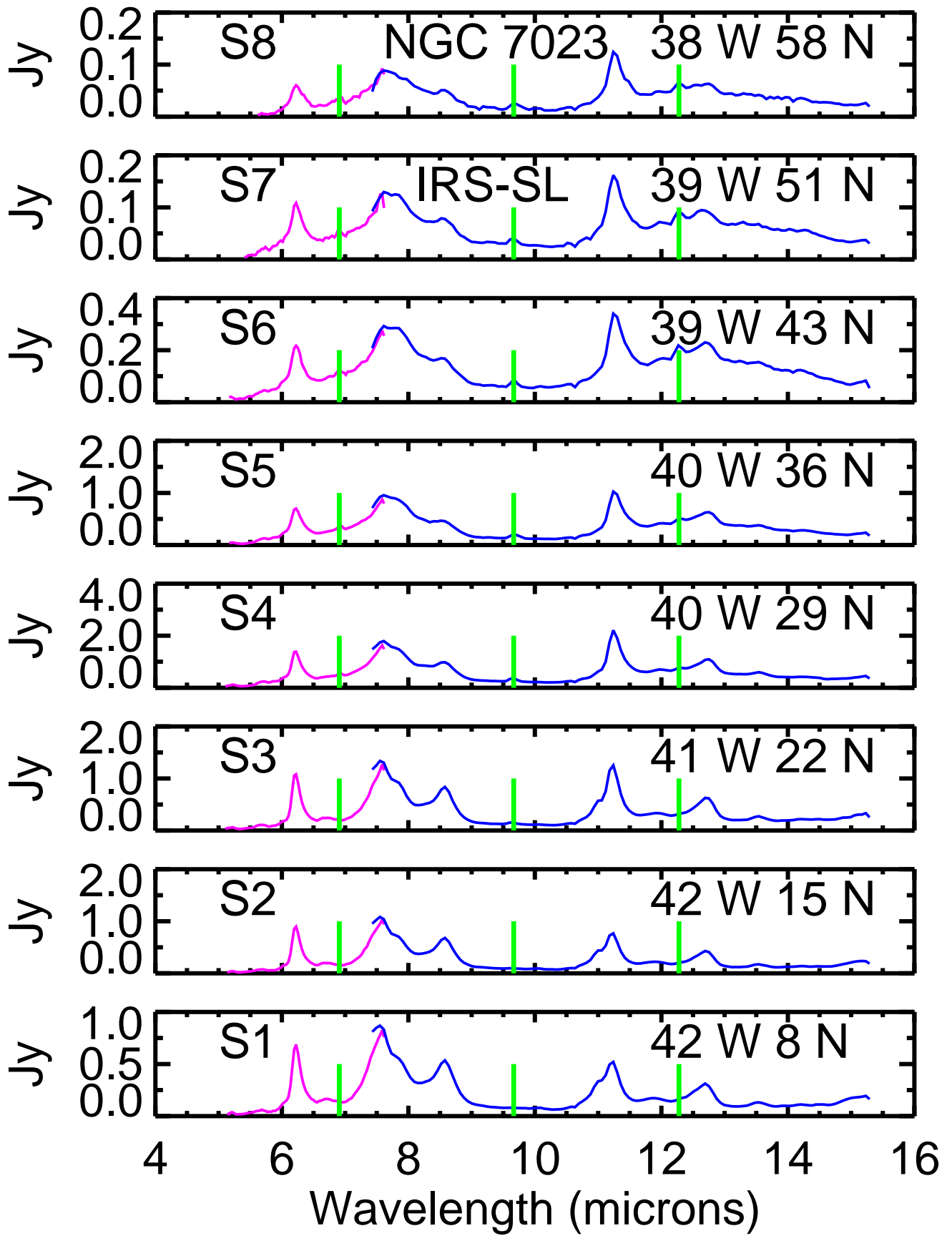
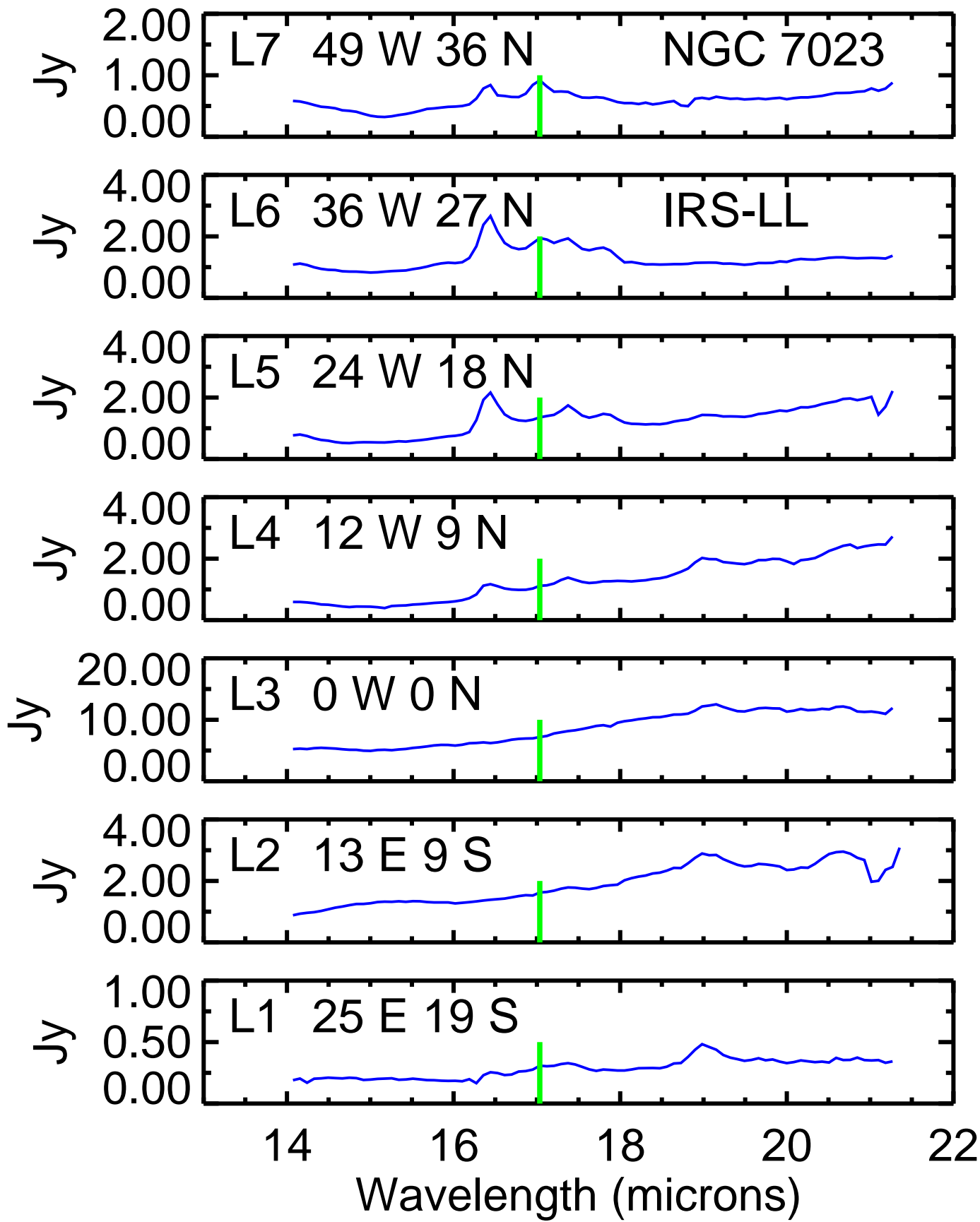
\end{document}